\documentclass[preprint2]{aastex}

\usepackage{graphicx,natbib}
\bibpunct{(}{)}{;}{a}{}{ }

\slugcomment{}

\shorttitle{Identifying black holes in {  X-ray binaries}}
\shortauthors{Barnard et al.}

\begin{document}


\title{ Black hole hunting in the Andromeda Galaxy}


\author{R. Barnard\altaffilmark{1}, J. P. Osborne\altaffilmark{2}, U. Kolb\altaffilmark{1} and C. A. Haswell\altaffilmark{1}}
 \altaffiltext{1}{The Open University, Milton Keynes, MK7 6AA, UK}
\email{R.Barnard@open.ac.uk}
\altaffiltext{2}{The University of Leicester, Leicester, LE1 7RH, UK }



\begin{abstract}
We present a new technique for identifying  stellar mass black holes
in low mass X-ray binaries (LMXBs), and apply it to XMM-Newton
observations of M31. We examine X-ray time series variability
seeking power density spectra (PDS) typical of LMXBs accreting at a low
accretion rate (which we refer to as Type A PDS); these are very similar
for  black hole and neutron star LMXBs. Galactic neutron star LMXBs
exhibit Type A  PDS at low luminosities (~10$^{36}$--10$^{37}$ erg/s) while black
hole LMXBs can exhibit them at luminosities $>$10$^{38}$ erg s$^{-1}$. We propose that
Type A PDS are confined to luminosities below a critical fraction of the
Eddington limit, $l_c$ that is constant for all LMXBs; we have examined a
sample of black hole and neutron star LMXBs and find they  are all
consistent with $l_c$ = 0.10$\pm$0.04  in the 0.3--10 keV band. We present luminosity and PDS data from 167 observations of X-ray binaries in M31 that provide strong support for our hypothesis.
Since the theoretical maximum mass for a neutron star is ~3.1 M$_{\odot}$,
we therefore assert that any  LMXB that exhibits a Type A PDS at  a
0.3--10 keV luminosity greater than 4$\times$10$^{37}$ erg s$^{-1}$ is likely to contain a
black hole primary. We have found eleven new black hole candidates in M31
using this method. We focus on XMM-Newton observations of RX J0042.4+4112,
an X-ray source in M31 and find the mass of the primary to be 7$\pm$2
M$_{\odot}$,  if our assumptions are correct. Furthermore, RX J0042.4+4112
is consistently bright in ~40 observations made over 23 years, and  is
likely to be a persistently bright LMXB; by contrast all known Galactic
black hole LMXBs are transient. Hence our method may be used to find black
holes in known, persistently bright Galactic LMXBs and also in LMXBs in
other galaxies.
 \end{abstract}





\section{Introduction}
\label{sec:intro}

Van der Klis (1994) showed that the power density spectra (PDS) of  low mass X-ray
binaries (LMXBs) with neutron star primaries at low  accretion rates are strikingly similar to
those of black hole LMXBs in their low   accretion rate states. The  fractional rms variability is high ($\sim$30--50\%) and the PDS  are well described  by a broken power law that changes in spectral index, $\gamma$, from $\sim$0 to $\sim$1 at frequencies higher than a certain break frequency; the break occurs at 0.01--1 Hz. 
 We will refer to these as Type A PDS. At higher accretion rates, the
rms variability is only a few percent, and the PDS are characterised by a power law with $\gamma$ $\sim$1--1.5. We will refer to these as Type B PDS.  Van der Klis (1994) proposed
that the transition between Type A and Type B PDS occurs when
the accretion rate exceeds a critical fraction of the Eddington limit ($f_{\rm c}$)
that is constant for all LMXBs. He   suggested $f_{\rm c}$ $\sim$1\%, as an order of magnitude estimate.

\begin{figure*}[!t]
\resizebox{\hsize}{!}{\includegraphics[angle=270]{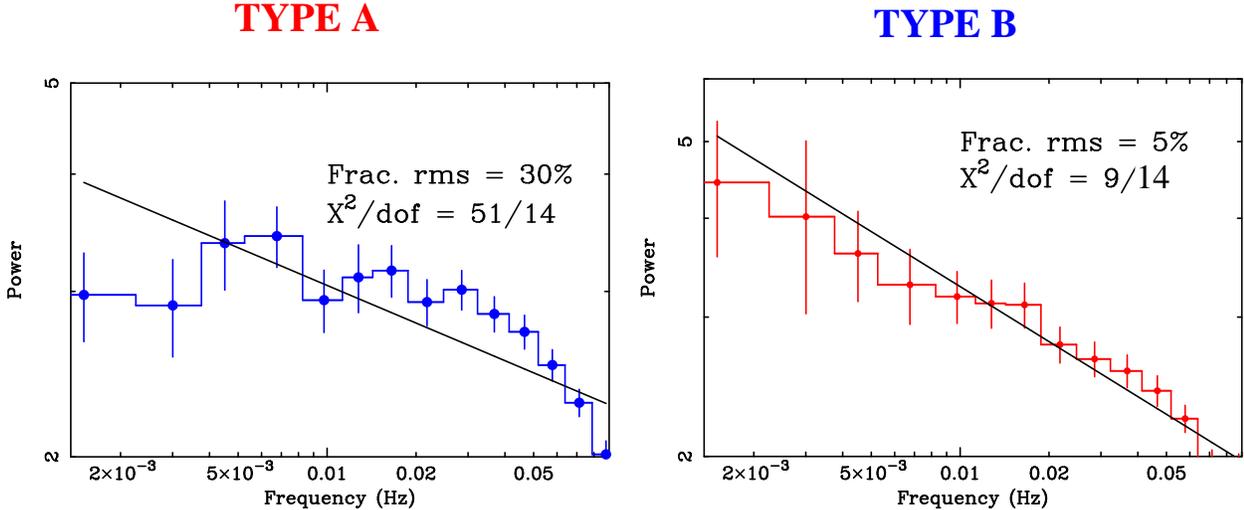}}
\caption{  Example PDS from XMM-Newton observations of the central region of M31; the axes are log-scaled and the PDS are Leahy normalised, so that Poisson noise has a power of 2. The left panel shows a Type A PDS: the best fit power law to the PDS has an unacceptable  $\chi^2$ of 51/14, and the fractional r.m.s. variability of the lightcurve is 30\%. The right panel shows a  Type B PDS: its shape is well described by a power law ($\chi^2$/dof = 9/14), and the fractional r.m.s variability of the lightcurve is 5\%.}\label{lhp}
\end{figure*}

We propose here  a new {  diagnostic} for identifying stellar-mass black holes in LMXBs.  Since we cannot observe $\dot{m}$ directly, we must use the lumosity to trace the evolution of the PDS with $\dot{m}$. 
We define $l$ as  $L/L_{\rm Edd}$, where $L$ is the luminosity and $L_{\rm Edd}$ is the Eddington limit, and $l_{\rm c}$ = $L_{\rm c}/L_{\rm Edd}$ so that Type A PDS are exhibited by LMXBs when  $l$ $<$ $l_{\rm c}$ and Type B PDS are exhibited when $l$ $>$ $l_{\rm c}$. Throughout this work we use the Eddington limit for uniform spherical accretion of hydrogen. However, the picture is obviously more complicated. Nevertheless, if the accretion discs of black hole and neutron star LMXBs are generally similar, then $l_{\rm c}$ should be similar also.

Initially, we assume that $l_{\rm c}$ is constant for all LMXBs, allowing black hole LMXBs to exhibit Type A  PDS at considerably higher luminosities than neutron star LMXBs. The Eddington limit   for hydrogen rich gas is $\sim$1.3$\times$10$^{38}$ ($M_{1}$/M$_{\rm \odot}$) erg s$^{-1}$, where $M_{1}$ is the mass of the primary; hence the maximum luminosity for a neutron star exhibiting a Type A PDS, $L_{\rm c}^{\rm NS}$ is determined by the  maximum mass of a neutron star. This allows us to classify the primaries of LMXBs that exhibit Type A PDS at luminosities higher than $L_{\rm c}^{\rm NS}$ as black hole candidates. We will assume a maximum neutron star mass of 3.1 $M_{\odot}$, as is generally accepted.

 Here,  we use XMM-Newton observations of globular cluster X-ray sources in M31 and results from the literature on Galactic LMXBs to find an empirical value of $l_{\rm c}$. We then apply our technique to four XMM-Newton observations of the 63 brightest X-ray sources in the central region of M31, before focusing on RX\thinspace J0042.4+4112, an X-ray source in the vicinity of M31 that   appears to exhibit both Type A and Type B PDS, allowing us to estimate its mass.

\section{Obtaining an empirical value for $l_{\rm c}$ in the 0.3--10 keV band}

In practise, we do not observe the full bolometric luminosity of an X-ray source and can only obtain its luminosity in a given energy band. We have been using the XMM-Newton and Chandra X-ray observatories to identify black hole LMXBs in external galaxies, and these have an energy range of $\sim$0.3--10 keV. We therefore estimate $l_{\rm c}$ in the 0.3--10 keV band, i.e. $L_{\rm c}^{\rm 0.3-10 keV}$/L$_{\rm Edd}$, since it is directly applicable to our observations. The energy spectra of neutron star and black hole LMXBs are similar at low accretion rate; hence $l_{\rm c}$ should scale to different energy bands in a similar way for all LMXBs.

 \subsection{  Estimating $l_{\rm c}$ from XMM-Newton observations of 14 globular cluster LMXBs in M31}

\begin{figure}[!t]
\resizebox{\hsize}{!}{\includegraphics[angle=270]{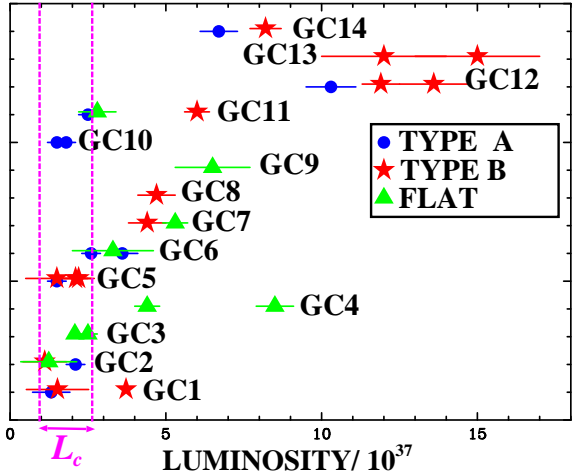}}
\caption{   The 0.3--10 keV luminosities of 14 globular cluster X-ray sources in  XMM-Newton observations of M31; the errors in the luminosity are given by 90\% errors in the parameters of the best fit spectral model. The dashed, vertical lines represent the range of $L_{\rm c}$, the luminosity of transition between Type A and Type B PDS for these sources   (presumed 1.4 M$_{\odot}$ neutron stars).  Bo 153 and Mita 299 are likely to be multiple bright X-ray sources, but are possible black holes.}\label{gcl}
\end{figure}

  Of the 63 X-ray source that we studied in the central region of M31,   14  were associated with globular clusters by \citet{K02}. There are thirteen  bright X-ray sources in Galactic globular clusters, and twelve of these  have been identified as neutron star LMXBs, while the thirteenth has not been classified \citep[][ and references within]{int04}. Hence the 14 globular cluster X-ray sources in our sample are expected to be LMXBs containing $\sim$1.4 M$_{\odot}$ neutron stars. However, note that \citet{ang01} have reported a possible globular cluster black hole binary in the elliptical galaxy \object{NGC\thinspace 1399}, with a 0.3--10 keV luminosity of 5$\times$10$^{39}$ erg s$^{-1}$.

Figure~\ref{gcl} shows the observed luminosities of the {  14} globular cluster X-ray sources for each of the  observations that were selected; a circle represents an observation where a Type A PDS was observed, { a star represents an observation where a Type B  PDS was observed and a triangle represents an observation where the PDS had no power, consistent with Type B, but not Type A; quoted uncertainties in the luminosity are based on 90\% confidence limits on the best fit parameters used to model the spectrum.

Each X-ray source is consistent with the hypothesis that Type A PDS are exhibited at lower luminosities than Type B PDS, { within errors}. {    Furthermore, twelve are consistent with 1.0 $\le$ ($L_{\rm c}$/ 10$^{37}$ erg s$^{-1}$) $\le$ 2.6}, assuming a distance {  to M31} of 760 kpc \citep{vdb00}; this corresponds to $l_{\rm c}$ = 0.10$\pm$0.04  if the primaries in these twelve sources are 1.4 M$_{\odot}$ neutron stars.
 Bo 153 and  Mita 299 exhibit   Type A {  PDS} at  luminosities of  (10.3$\pm$0.8) and (6.7$\pm$0.6) $\times$10$^{37}$ erg s$^{-1}$ respectively.   These systems would be consistent with our hypothesis if they contained  black hole binaries, or if they were  composed of two or more bright X-ray sources \citep[like M15, see][]{wa01} or multiple faint X-ray sources, in the globular clusters \citep[see e.g.][]{hgl03}.  
}

\subsection{ $l_{\rm c}$ for a Galactic neutron star LMXB}
4U\thinspace 1705-44 is a Galactic LMXB that exhibits X-ray bursts, and hence contains a neutron star \citep{lang87}. It exhibited a Type A  PDS in the faintest of four EXOSAT observations, and a Type B PDS in the next faintest; the respective 1--11 keV fluxes were 1.3$\times$10$^{-9}$ and 1.8$\times$10$^{-9}$ erg cm$^{-2}$ s$^{-1}$ \citep{lang87,lang89}. Hence, an accurate distance would yield a tight constraint on $l_{\rm c}$. The distance to 4U\thinspace 1705$-$44 has been estimated using X-ray bursts as standard candles \citep[see][]{kul03};  \citet{cs97} find a distance of 11 kpc from Einstein data, while \citet{cki02} obtain a distance of 8.6 kpc using data from BeppoSAX. If we assume that the distance lies between these two values, $l_{\rm c}$ {  = } 0.10$\pm$ 0.04.

\subsection{ $l_{\rm c}$ for  Galactic black hole LMXBs}

\begin{figure}[!t]
\resizebox{\hsize}{!}{\includegraphics[angle=270]{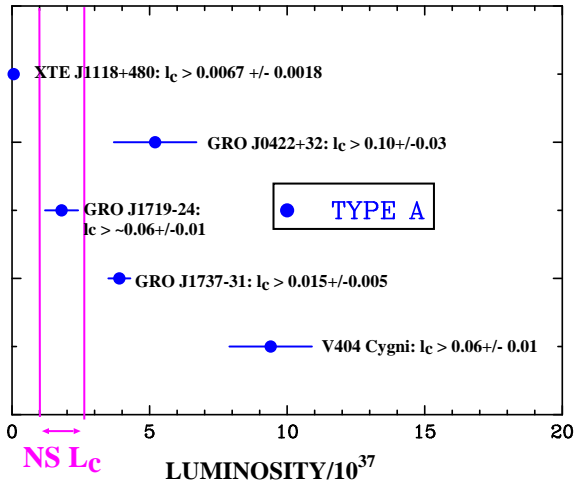}}
\caption{  Luminosities of Type A PDS from 5 Galactic black hole LMXBs; lower limits to $l_{\rm c}$ are given for each object.  }\label{gbh}
\end{figure}

Of the eighteen confirmed Galactic black hole X-ray binaries, three are high mass X-ray binaries (HMXBs), and are persistently bright, and the rest are transient LMXBs.  In general  outbursts last several months and the X-ray luminosity can increase by a factor of 10$^7$; {  outbursts are} on average separated by years of quiescence \citep{cst97,int04}.  The outbursts are hysteretic in that the transition from the low/hard state to the high/soft state during the rise of the outburst occurs at a higher luminosity than transition from the high/soft state to the low/hard state in decay \citep[e.g.][]{miy95,mac03}; hence, estimates of $l_{\rm c}$ in black hole LMXBs were restricted to those that have been observed during the rise of the outburst, and the subset of outbursts where the transition from low/hard state to high/soft state were not made \citep[see][ for a review]{bbf04}. Unfortunately, most X-ray observations of Galactic black hole LMXBs during outburst have been during the decay phase, and are hence unsuitable for this work.

Two of the    confirmed black hole LMXBs,   GX\thinspace 339$-$4 \citep{zdz04} and  XTE\thinspace J1550$-$564 \citep{rod03}, have recently been caught during the rise of the out burst by the RXTE-ASM, allowing monitoring of the entire outburst by the main instruments of RXTE. Both systems exhibited spectral transitions at bolometric luminosities of $\sim$20\%. This corresponds to 0.3--10 keV luminosities of $\sim$10\% Eddington, i.e. $l_{\rm c}$ $\sim$0.1 in the 0.3--10 keV band.

Additionally, nine  black hole LMXBs  have exhibited outbursts where they remained in the low/hard state \citep[for a review see][]{bbf04}. Five of these have published {  Type A} PDS, distances and mass estimates; we have used published results to obtain the {  corresponding} minimum value of $l_{\rm c}$ in the 0.3--10 keV band. The results are presented in Fig.~\ref{gbh}; the mass of GRS\thinspace 1737$-$31 is given by the mass range of known black holes. {  We note that    \citet{es98} find that the SED of   GRO\thinspace 0422+32 at the peak of its outburst is consistent with an accretion rate of $\sim \dot{m}_{\rm crit}$; this is interesting because we find  the 0.3--10 keV luminosity of GRO\thinspace 0422+32 at that time to be 10$\pm$3\% Eddington. Results from the other four black hole LMXBs are also consistent with  $l_{\rm c}$ $\sim$0.1 in the 0.3--10 keV  band, although they cannot constrain $l_{\rm c}$.

\subsection{Our empirical value of $l_{\rm c}$}

The data are all consistent with $l_{\rm c}$ in the 0.3--10 keV band of 0.10$\pm$0.04 for neutron star and black hole  LMXBs.   Using $L_{\rm c}^{\rm NS}$ = 3.1 M$_{\odot}$, we classify LMXBs that exhibit Type A PDS at $>$4$\times$10$^{37}$ erg s$^{-1}$ as likely black hole LMXBs. So far we have identified 11 candidates in the core of M31, including RX\thinspace J0042.3+115 \citep{bok03} and CXOM31\thinspace J004303.2+411528 \citep*{bko04}.

{
\section{The bright X-ray population of the central region of M31}

\begin{figure}[!t]
\resizebox{\hsize}{!}{\includegraphics[angle=270]{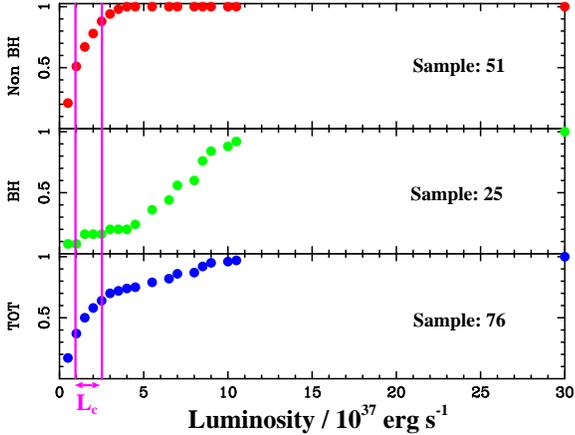}}
\caption{ Cumulative distribution function (CDF) vs. 0.3--10 keV luminosity for Type A PDS for the brightest X-ray sources in four XMM-Newton observations of the central region of M31. The bottom panel shows the CDF for the whole sample, and the sample is divided into black hole candidates and non-black hole candidate in the middle and top pannels respectively. The vertical lines represent the reange of $L_{\rm c }$ for a 1.4 M$_{\odot}$, derived from Fig.~\ref{gcl}. The most striking feature is the break in the CDF of the total sample, at $\sim$1.5--3$\times$10$^{37}$ erg s${-1}$; this is what we expected for an X-ray population dominated by LMXBs with a 1.4 M$_{\odot}$ neutron star primary, if $l_{\rm c}$ has a constant value of $\sim$0.1. }\label{cdfa}
\end{figure}

\begin{figure}[!t]
\resizebox{\hsize}{!}{\includegraphics[angle=270]{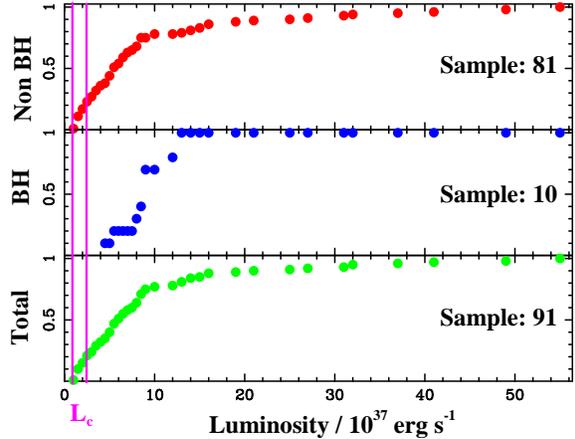}}
\caption{ Cumulative distribution function vs. 0.3--10 keV luminosity for Type B and flat (cannot be Type A, probably Type B) PDS, analogous to Fig.~\ref{cdfa}. We note that no Type B PDS are observed in the black hole sample below 4$\times$10$^{37}$ erg s$^{-1}$.  }\label{cdfb}
\end{figure}

The 0.3--10 keV luminosities and PDS of the 63 brightest X-ray sources were analysed for each of the four XMM-Newton observations of the central region of M31. The sample is likely to be dominated by X-ray binaries, and since the field of view is dominated by the bulge, they are most likely to be LMXBs.  The sample was selected on the criterion that the average 0.3--10 keV EPIC-pn intensity was greater than 0.02 count s$^{-1}$ in at least one of the four observations. For each observation of each source, a PDS was made of the combined EPIC, background subtracted, 0.3--10 keV lightcurve; each PDS was averaged over many intervals, which were  divided into 128 bins of 5.2 s duration. Luminosities were obtained from the unabsorbed 0.3--10 keV flux given by best fit models to EPIC-pn spectra.

From this sample, all known foreground objects and background AGN identified by \citet{K02} were filtered out, as were sources  that were resolved into multiple X-ray sources by Chandra. Furthermore, any observation where the classification of the PDS was ambiguous was also rejected. 

Of the 167 observations accepted, 76 exhibited Type A PDS and 91 exhibited Type B or flat  PDS. X-ray sources that exhibited Type A PDS at luminosities higher than 4$\times$10$^{37}$ erg s$^{-1}$ in at least one of the four observations were classed as black hole candidates, the rest were classed as non-black holes. We note that these black hole identifications are neither certain nor complete. 

Cummulative distribution function (CDFs) of Type A PDS vs luminosity are  presented in Fig.~\ref{cdfa}; the CDFs for the non-black hole population is shown in the top panel; the black hole CDF is shown in the middle panel, and the CDF for the whole sample is shown in the bottom panel. The vertical lines indicate the range of transition $L_{\rm c}$ obtained from the globular clusters in Fig.~1, assuming 1.4 M$_{\odot}$ neutron star primaries. We see that the total CDF exhibits a natural break in the region 1.5--3$\times$10$^{37}$ erg s$^{-1}$, which is exactly what we would expect for a population dominated by 1.4 M$_{\odot}$ neutron star LMXBs, if $l_{\rm c}$ $\sim$ 0.1. Also, 50\% of the 51 non-BH Type A PDS were seen at luminosities below 1.0$\times$10$^{37}$ erg s$^{-1}$, despite the expected strong bias towards high luminosity Type A PDS due to improved statistics.

The  CDFs vs. luminosity of Type B and flat  PDS for the non-black hole, black hole and total populations of X-ray sources are presented in the top, middle and bottom panels of  Fig.~\ref{cdfb} respectively.  As expected, the majority ($\sim$80\%) are observed at luminosities higher than the range of $L_{\rm c}$ values obtained from the globular cluster population, and none are observed below the lower limit to $L_{\rm c}$; a break occurs around 10$^{38}$ erg s$^{-1}$, further indicating that most of the sources contain neutron stars.We do not expect all Type B PDS to occur above $l_{\rm c}$, since some of the sources are likely to be transient, and hence hysteretic. Most significantly, none of our black hole candidates exhibit Type B PDS below 4$\times$10$^{37}$ erg s$^{-1}$; this provides strong support for the idea that Type B PDS are observed at higher luminosities than Type A PDS, and suggests that these black hole candidates are not hysteretic, and are perhaps persistently bright. We note however, that only 10 Type B PDS were observed in our black hole candidates.

\section{RX\thinspace J0042.4+4112}

\begin{figure}[!t]
\resizebox{\hsize}{!}{\includegraphics[angle=270]{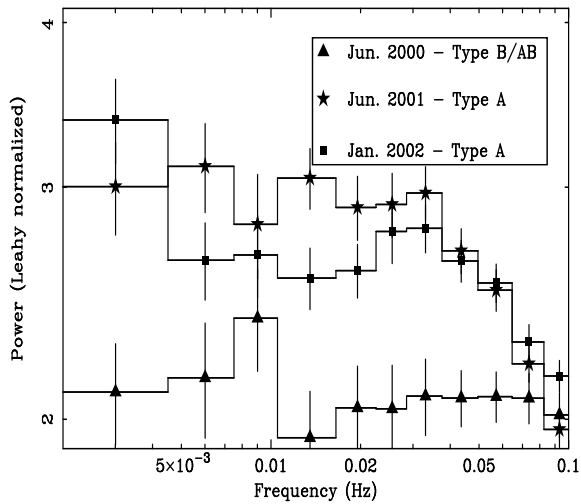}}
\caption{Power density spectra of 0.3--10 keV lightcurves from XMM-Newton observations of RX\thinspace J0042.4+4112. The PDS are Leahy normalized, so that the Poisson noise has a power of 2.  }\label{pds}
\end{figure}

RX\thinspace J0042.4+4112 is located at  00h42m28$\fs$268 +41$\degr$12$\arcmin$22${\farcs}$76 \citep{K02} and is particularly exciting because,  it is a black hole LMXB by our classification that also exhibited a flat (i.e. not Type A) PDS, allowing us to estimate a primary mass.}
The PDS of combined EPIC (MOS1 + MOS2 + PN) 0.3--10 keV  lightcurves  from three XMM-Newton observations of RX\thinspace J0042.4+4112 are presented in Fig.~\ref{pds}. The 0.3--10 keV SEDs were well described by an absorbed power law with spectral index $\Gamma$ $\sim$1.7--1.8.
 Table~\ref{lpds} gives $\Gamma$, the luminosity and PDS type for each observation.

\begin{table}[!t]
\caption{\sc The spectral index ($\alpha$), luminosity and PDS Type of RX\thinspace J0042.4+4112 in each of three XMM-Newton observations.}\label{lpds}
\centering
\begin{tabular}{ccccc}
\noalign{\smallskip}
\tableline
\tableline
\noalign{\smallskip}
Observation& $\Gamma$ & $L$\tablenotemark{a} &         PDS Type\\
\noalign{\smallskip}
\tableline
\noalign{\smallskip}
2000 Jun 25 & 1.71$\pm$0.07& 9.1$\pm$0.6 & FLAT (B?)\\
2001 Jun 9 &1.74$\pm$0.06 & 8.0$\pm$0.5 & A \\
2002 Jan 6 & 1.79$\pm$0.05 & 8.4$\pm$0.5 & A \\
\noalign{\smallskip}
\tableline
\end{tabular}
\tablenotetext{a}{0.3--10 keV luminosity / 10$^{37}$ erg s$^{-1}$}
\end{table}

RX\thinspace J0042.4+4112 clearly exhibits Type A PDS in the 2001 and 2002 observations. However, no variability is detected in the 2000 observation, despite the higher  luminosity. Since we see Type A PDS in the 2001 and 2002 observations, the lack of variability and higher luminosity in the 2000 observation is consistent with a Type B PDS. This would imply that the $l$ $>$ $l_{\rm c}$ in the 2000 observation.
 If the timing states of RX\thinspace J0042.4+4112  indeed resemble those of Galactic  LMXBs, then we can estimate the mass of the primary.

   We find that $L_{\rm c}$ = 8.8$\pm$0.9$\times$10$^{37}$ erg s$^{-1}$; assuming that $l_{\rm c}$ = 0.10 $\pm$0.04, the  primary mass is 7$\pm$2 M$_{\odot}$, which is above the maximum mass of a neutron star

RX\thinspace J0042.4+4112 is consistently bright in the 4 XMM-Newton observations, as well as  in the 1991 and 1992 ROSAT surveys of M31 \citep{S97,S01},  the 1979 Einstein observation \citep{tf91}, and in all 30 Chandra observations of the source made between October 1999 and September 2002. Hence RX\thinspace J0042.4+4112 is likely to have been persistently bright for over 20 years. 

\section{Acknowledgments}
This work was funded by PPARC. R.B. would like to thank W. Clarkson, R. Cornelisse, T. Maccarone, J. Wilms and D. Gelino for useful discussions. 

\bibliographystyle{aa}
\bibliography{m31}

\end{document}